\documentclass[superscriptaddress,twocolumn,pre]{revtex4}
\usepackage{amssymb, amsmath, graphicx}
\def\be{\begin{equation}}
\def\ee{\end{equation}}
\begin{document}
\title{Non-stationary Markovian Replication Process causing Diverse Diffusions}
\author{Yichul Choi}
\affiliation{
Department of Physics, Virginia Polytechnic Institute and State University, Blacksburg, 
Virginia 24061-0435, USA}
\author{Hyun-Joo \surname{Kim}}\email{hjkim21@knue.ac.kr}
\affiliation{
Department of Physics Education, Korea National University of Education,
Chungbuk 363-791, Korea}

\received{\today}

\begin{abstract} 
We introduce a single generative mechanism with which it is
able to describe diverse non-stationary diffusions. A non-stationary Markovian replication process for steps
is considered, for which we analytically derive time-evolution of the probability distribution of the walker's displacement 
and the generalized telegrapher equation with time-varying coefficients, and find that diffusivity can be determined 
by temporal changes of replication of a immediate step. 
By controlling the replications,  we realize the diverse diffusions such as  alternating diffusions, superdiffusions, subdiffusions,
and marginal diffusions which are originated from oscillating, increasing,
decreasing, and slowly increasing or decreasing replications with time, respectively.

\end{abstract}

\maketitle
\section{Introduction}
Starting with the purpose of understanding the random motion of Brownian particles,  diffusive phenomena 
have been received great attention for a long time in the statistical physics  as well as in recent various fields 
such as human geographical \cite{hm,hm1,hm2,hm3}, hydrological \cite{river,hjkim}, 
biophysical \cite{bio1,bio2,bio3,bio4,bio5,bio6}, 
economic systems\cite{econo1,econo2}, and so on. The Brownian motion 
follows the Fokker-Planck  equation (FPE) well known as the diffusion equation for the  probability density
function (PDF) from which the mean-squared displacement (MSD) is linearly dependent
on time, $\langle x_t ^2 \rangle = 2D_0 t$  where $D_0$ is the constant diffusion coefficient. 
This Brownian process  is well described by  a stationary Markovian model known as random walk \cite{rw,ks}.
However recent studies
report that MSD shows the nonlinear behavior rather than the linear behavior for time  \cite{godec,the1,the2,swarm,exp1,exp2}. 
The MSD following the power-law behavior,  $\langle x^2 (t) \rangle \sim t^{2H}$ characterizes
anomalous diffusion, where  $H$ is called as the Hurst exponent which classifies superdiffusion
($ H > 1/2 $) in which the past and future random variables are positively correlated and thus 
persistence is exhibited, and subdiffusion ($ 0 < H < 1/2 $) which behaves in the opposite way,
showing antipersistence. 

Efforts to describe mechanisms underlying anomalous diffusions have been tried through
representative stochastic models such as fractional Brownian motion (fBM)
where long-ranged temporal correlation between steps is given and the Hurst exponent ranges from 0
to 1 \cite{fbm}, L\'{e}vy walk model that describes well superdiffusions by drawing a step length from the distribution with a heavy
power-law tail and keeping a constant speed for a random time \cite{levir,lrtlw,leviwalk}, 
continuous time random walks (CTRW) with the power-law distribution of time intervals for a step showing subdiffusions \cite{rw,ctrw},
and scaled Brownian motion(sBM) which is described by a diffusion equation with explicitly 
time-dependent diffusion coefficient \cite{sbm,sbm1}. 
These models show the non-stationarity or non-Markovianity which are responsible for the anomalous
diffusive behaviors. The  fBM is non-Markovian but stationary, and the L\'{e}vy walk, the CTRW and the sBM 
are semi-Markovian but non-stationary. 

In addition, the stochastic models with the memory of whole previous trajectory in a walk process 
mimicking the movements of animals such as elephants \cite{pre70} and monkeys \cite{boyer},
has been introduced
and it is known that memorizing the history of a process which make a process be non-Markovian and non-stationary
plays a key role in generating the long-term correlations between steps resulting in anomalous diffusions
\cite{pre70,boyer, mm1,mm2}. 
However, memorizing whole history of previous steps is not easy and plausible except for
some specific cases, rather it is much more acceptable to consider short-term memory like
remembering just the immediate step. 
Although it did not start from the perspective of the short-term memory,
it was already considered in the 
persistent random walk model \cite{goldstein}  in which a step follows the previous step with a constant probability,
resulting in a movement to the same direction that the walker was moving.
Also it is known that such a process does not follow the diffusion equation but the telegrapher equation (TE) \cite{tele}
which has an additional second order time derivative term of the PDF to the diffusion equation which introduces 
wave equation property and is related to ballistic motion of the diffusion particles, but   
asymptotically reduces to normal diffusive behavior \cite{goldstein,ftele}. That is, although the telegrapher process has
advantages in describing ballistic motion in early stages and is applicable to the diverse diffusion and 
transport phenomena \cite{ftele, tele1,tele2}, eventually, 
it is a stationary normal diffusive process and not sufficient to explain property of nonstationary 
movements resulting in diverse diffusions appeared in nature .

The nonstationary movements of living organisms are natural in making a adaptation for the various types of
temporal stimuli coming from their natural environments\cite{ecoli}. 
In particular, in kinetics of eukaryotic cells under temporal chemotatic or mechanotactic signaling,
it has been studied that cells response directly by changing their motion depending on temporal stimuli
\cite{ts1,ts2,ts3}.
That is, to respond to such complex temporal stimuli a walker may move to the opposite direction to a previous step in a momemt,
or conversely,  strengthen movements to the same direction. Thus memorizing previous steps can change with time
and we
modelize it with a time-varying replication probability which controls the degree of following the immediate step at the next step.
Namely, a non-stationary persistent random walk model is introduced and a generalized TE 
with time-dependent coefficients is derived. We also calculate the relations between the time-dependent coefficients and the replication
probability and thus show that explicit time dependence of the probability could produces long-term correlation between steps
which results in diverse diffusions deviated from a normal diffusive behavior.

\section{Non-stationary Markovian replication model}
We consider a walker on a one dimensional homogeneous lattice with the uniform spacing  $l$ between the
neighboring sites. With a regular time interval denoted as $\tau$, 
the walker moves to one of the two neighboring sites. 
We denote the walker's position at time $t$ as $x_t$, and the step walker takes as
$\sigma_t$, which is defined by the relation 
\be
x_{t}=x_{t-\tau}+\sigma_{t}.
\label{xs}
\ee
The walker is initially at the origin.
Details of the model is determined by a time-varying probability $\alpha(t)$ which controls
the dynamics of the process, 
\be
\sigma_{t} = \left \{
\begin{array}{ll}
\sigma_{t-\tau},  & \quad \text{with a probability $\alpha(t)$}\\
-\sigma_{t-\tau},  & \quad \text{with a probability $1-\alpha(t)$.}
\end{array} \right.
\label{erule}
\ee
The first step($\sigma_\tau$) is randomly chosen between the two possibilities $\pm l$
with the equal probabilities $1/2$.
Successive steps and positions at $t \geq 2\tau$ are determined by Eq.~(\ref{xs}) and Eq. (\ref{erule}). 
We note that the process defined as above is symmetric about the origin.

At each time $t$, a step $\sigma_t$ replicates or anti-replicates the latest step $\sigma_{t-\tau}$.
Since the next step is completely determined by the immediate step, this replication process is Markovian, 
while the probability of replication, $\alpha(t)$, varies with time in general. Even if such a non-stationary nature is present, 
Markovianity of the step process makes the process analytically tractable.
In terms of kinematics, anti-replication of the latest step corresponds to change of direction of motion, 
and when $\alpha(t)$ is constant, the model reduces to the persistent random walk model \cite{goldstein}.
From the perspective of memorizing a trajectory
and concerning possible applications of the model not only to the diffusion processes in real space but also 
to the analysis of general two states time series, we prefer looking Eq. (\ref{erule}) as a replication-antireplication 
process rather than just alternating direction of motion. Thus, we call $\alpha(t)$ and the process~(\ref{erule}) 
the replication probability and the non-stationary Markovian replication process (NMRP), respectively.

\section{Time Evolution of the PDF and the MSD}
Now, we derive the time evolution of the probability distribution of the displacement $P(x,t)$ for the NMRP model, which
is the probability that the walker's position is $x$ at time $t$, 
starting from the relation
\be
P(x,t)=\sum_{x_{t-\tau},x_{t-2\tau}}^{} P(x,t|x_{t-\tau},x_{t-2\tau})P(x_{t-\tau},x_{t-2\tau}).
\label{mestart}
\ee
Here, $P(x_{t-\tau},x_{t-2\tau})$ is the probability that the walker's positions
at times $t-\tau$ and $t-2\tau$ are $x_{t-\tau}$ and $x_{t-2\tau}$, respectively, and $P(x,t|x_{t-\tau},x_{t-2\tau})$ is
the second order transition probability, which is a conditional probability that
the walker's position is $x$ at time $t$ given the two previous positions. 
The summation runs over all lattice sites.
Because of the Markovianity of the step process (\ref{erule}),  
it is possible to use the second order transition probability which 
is determined in terms of only $\sigma_{t}$ and $\sigma_{t-\tau}$ at time $t$ and  is expressed by
\be
\begin{split}
P(x,t|x_{t-\tau},x_{t-2\tau})=
\{ \delta_{\sigma_{t-\tau},l}+\delta_{\sigma_{t-\tau},-l}\}  \\
\times \{ \alpha(t)\delta_{\sigma_{t},\sigma_{t-\tau}}
+[1-\alpha(t)]\delta_{\sigma_{t},-\sigma_{t-\tau}}\}. 
\end{split}
\label{transition}
\ee
Terms in the first bracket depict the two possible choices for the step $\sigma_{t-\tau}$, and
the other two terms in the second bracket indicate the probabilities that the replication or anti-replication occur at time $t$, respectively. 

By substituting the Eq. (\ref{transition}) into the Eq. (\ref{mestart}),
the time evolution of $P(x,t)$ is described by the following master equation,
\be
\begin{split}
P(x,t)=&\alpha(t) \{P(x+l,t-\tau)+P(x-l,t-\tau) \} \\
& + \{ 1 - 2\alpha(t) \} P(x,t-2\tau),
\end{split}
\label{me}
\ee
which is valid for $t \geq 2\tau$.
If $\alpha(t)={1 / 2}$, the last term on the right hand side
vanishes, and Eq.~(\ref{me}) reduces to that of the normal random walk with symmetric probabilities.

Next, we take the continuum limit, considering the position and the time approximately as continuous variables. Expanding the master equation Eq. (\ref{me}) into a Taylor series keeping the lowest non-vanishing order terms, the time evolution of the PDF $\rho(x,t)$ in the continuum limit is obtained as follows,
\be
{{\partial \rho(x,t)} \over {\partial t}} + \mathcal{R}(t) {{\partial ^ 2 \rho(x,t)} \over {\partial t^2}}\\
= \mathcal{D}(t) {{\partial ^ 2 \rho(x,t)} \over {\partial x^2}},
\label{fpe}
\ee
where
\be
\mathcal{R}(t)={\tau \over 2} \left[ {{3\alpha(t) -2} \over {1-\alpha(t)}} \right]
\label{rt}
\ee
and
\be
\mathcal{D}(t)= D_0 \left[ {{\alpha(t)} \over {1-\alpha(t)}} \right]
\label{dt}
\ee
with $D_0=l^2 / {2\tau}$ being the diffusion coefficient for the normal diffusion. Eq. (\ref{fpe}) becomes  a generalized TE with the 
persistent coefficient $\mathcal{R}(t)$ and the diffusion coefficient $\mathcal{D}(t)$ depending  on time.
The relation between the coefficients $\mathcal{R}(t)$ and $\mathcal{D}(t)$ and the replication probability $\alpha (t)$ are given 
by the Eq. (\ref{rt}) and  the Eq. (\ref{dt}), respectively, which indicates that the larger $\alpha (t) $, the larger coefficients.
Note that when $\alpha(t)$ approaches to 1, two coefficients $\mathcal{R}(t)$ and $\mathcal{D}(t)$ diverge, whereas the coefficient of the first term on the left hand side of Eq. (\ref{fpe}) remains constant. Therefore, if the divergence is fast enough, a contribution from the first term on the left hand side of Eq. (\ref{fpe}) becomes negligible.
In this case, Eq. (\ref{fpe}) evolves into the wave equation with the speed $v=l/\tau$, which implies the occurrence of the ballistic motion of the walker. In the 
telegraph process, initial ballistic motion is transient as mentioned already. However, in the NMRP, ballistic motion appears whenever the replication is more dominant than the anti-replication, namely, $\alpha(t)$ becomes close to 1.
In more usual cases where $\alpha(t)$ does not tend to 1 so fast, because of the relatively small value of the $\tau$ in the continuum limit and the asymptotically small nature of the second order derivative of the PDF in $t$ compared to the first order one, the second term on the left hand side of Eq. (\ref{fpe}) can be neglected. In this case, Eq. (\ref{fpe}) reduces to the diffusion equation for the PDF $\rho(x,t)$ with the time dependent diffusion coefficient $D(t)$ \cite{sbm,sbm1,boyer},
\be
{{\partial \rho(x,t)} \over {\partial t}} = D(t) {{\partial ^ 2 \rho(x,t)} \over {\partial x^2}}
\label{efk}
\ee
and the solution is given by the Gaussian distribution provided zero mean of displacement, 
\be 
\rho(x, t) = {1 \over {\sqrt {2 \pi \langle x_t ^2 \rangle}}} \text {exp} \left[ -{ {x^2} \over { 2 \langle x_t ^2 \rangle }} \right].
\label{gaussian}
\ee

Now, we show that there is a unique relation between the MSD of NMRP and the replication probability, and by
manipulating the replication probability, nearly any form of the MSD can be generated if the MSD does not exceed the ballistic limit set by the finite 
and constant maximum speed of the process, $v=l/\tau$.
By multiplying $x^2$ to both side of the Eq. (\ref{fpe}) and then integrating with respect to $x$ over all space,
the following relation between $\alpha(t)$ and the MSD is obtained,
\be
\alpha(t)={ {\dot{\langle x^2_t \rangle} - \tau \ddot{\langle x^2_t \rangle}} 
\over {{{l^2} \over \tau }+  \dot{\langle x^2_t \rangle}-{3\tau \over {2}}\ddot{\langle x^2_t \rangle}} }.
\label{alpha}
\ee
When the Eq. (\ref{efk}) could be considered, $\ddot{\langle x^2_t \rangle}$ in Eq. (\ref{alpha}) is removed 
and the MSD is calculated for general $\alpha(t)$ as
\be
\langle x^2_t \rangle = 2 D_0 \int_{}^{t} {\alpha(s) \over {1-\alpha(s)}  } ds.
\label{msd}
\ee
If  $\alpha(t)$ does not change in time, the process reduces to normal diffusion.
Thus, a stationary replication process can not make a deviation from the normal diffusive behavior in the asymptotic limit.
The time-varying property in the replication process becomes the key point inducing diverse diffusions. 

\section{Diffusions using several specific replication probabilities}

\subsection{Alternating diffusions}

In the experiments for the cellular motion, external temporal stimuli have been simply imposed by a
step-function change in chemo-effector concentration \cite{ts1}  or mechanotactic signaling of
repeating step-like type  \cite{ts2} and exponentiated sine-waves for more complex fluctuating signaling 
in time \cite{ts3}. For the responding movements to periodic temporal stimuli we can consider
the periodic replication probability and if the value of probability changes from 0 to 1, we can also study the  
motion alternating from totally anti-persistent phase to persistent phase. 
As such an example, we have chosen the replication probability of $\alpha(t)=\text{sin}^2 (\pi t/T)$, with the period $T=N/5$ 
where $N$ is the total number of steps.
 \begin{figure}[ht]
 \includegraphics[width=9cm]{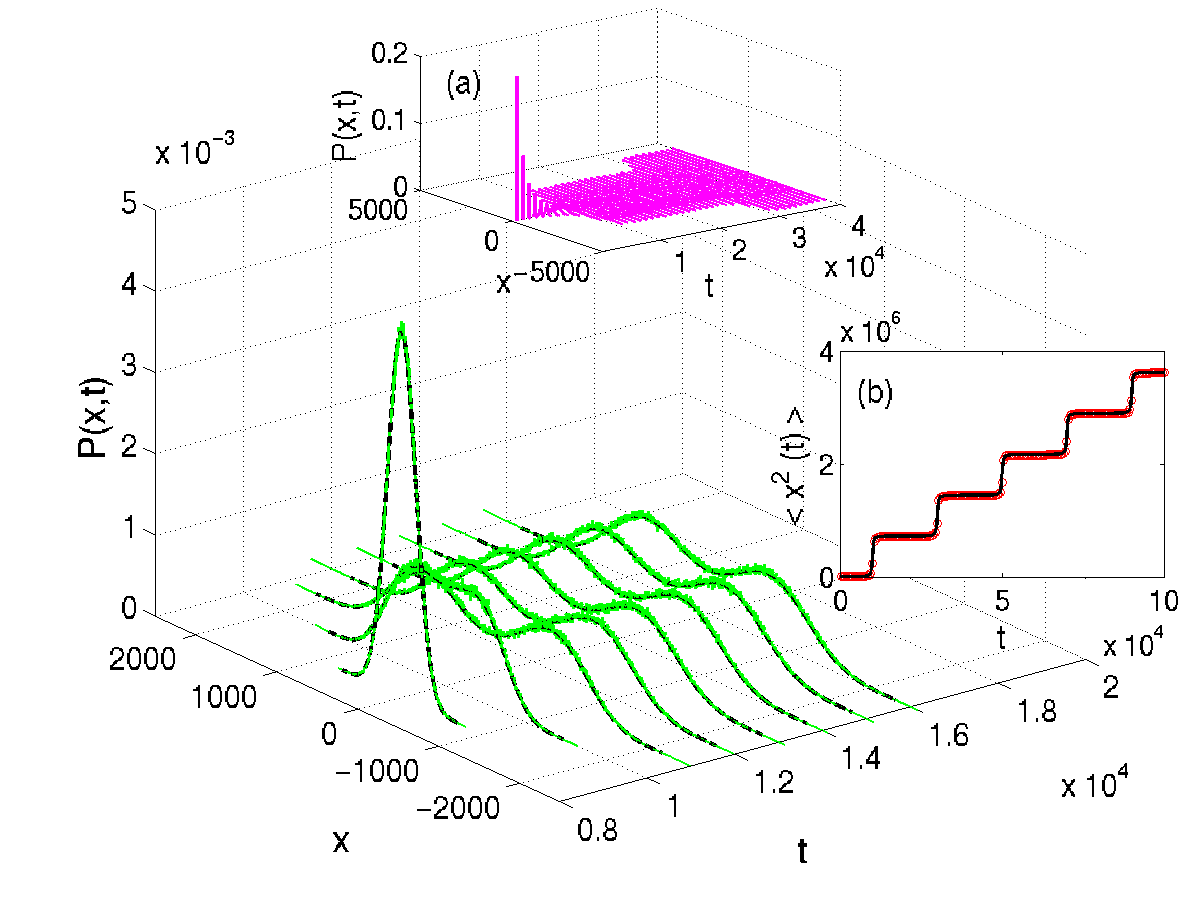}
 \caption{\label{fig1} Time evolution of $P(x,t)$ for the oscillatory replication process, $\alpha(t)=\text{sin}^2 (\pi t/T)$,
with the period $T=2 \times 10^4 \tau$. For the time interval between $0.4T$ and $0.8T$, the exact $P(x,t)$ obtained
from solving Eq. (\ref{me}) numerically (the black dashed lines) shows the perfect coincidence with the data  obtained 
by simulating the model (the green lines).
The inset (a) shows the simulation result of $P(x,t)$ for two periods,
in which periodic swelling and freezing of $P(x,t)$ is observed. The inset (b) shows the 
simulated (the red circles) and numerically solved (the black solid line) MSD which shows  a stair-like shape 
having periodic plateaus and sudden jumps. In this study, all simulations for the models have been done with the fixed 
values $\tau=l=1$.
 }
 \end{figure}

Overall, $P(x,t)$ is composed of periodically repeating swelling regions in which
$\alpha(t)$ is around the maximum value and thus 
almost perfect replication happens, and  freezing regions in which $\alpha(t)$ deviates from the 
maximum and the nature of anti-persistent is realized,  (the inset (a) of Fig. \ref{fig1}).
$P(x,t)$ around $\alpha(t)=1$ is enlarged in the main panel of Fig. \ref{fig1}
in which a single peak of $P(x,t)$ at the origin at $t=0.4T$ splits into two peaks away from the center
after $t=0.6T$. 
It shows that the walkers around the center is divided to the two opposite directions due to the almost perfect replications 
around $\alpha(t)=1$ and then decreasing $\alpha(t)$ results in freezing walkers and the peaks are maintained
until the next swelling point. However, after a several period such peaks
disappear because repeating of the perfect replications makes
much more small peaks and then  a peak at the center is restored.

Characteristics of the periodic oscillation in the replication process directly propagate 
into the probability and the MSD shows a interesting behavior with the periodically stair-like shape
in which plateaus of the MSD corresponds to the freezing regions of $P(x,t)$,
while sudden jumps appear in the swelling regions of $P(x,t)$ (the inset (b) of Fig. \ref{fig1}).
Such an stepwise increasing MSD has been reported in \cite{step} 
where there are two alternating waiting time distributions, one of which centered around zero and the other centered around 
some finite waiting time, where the standard deviations of both distributions are small. 
Such a setup leads to a movement in which the walker periodically repeats two sudden jumps and waitings where the waiting time is nearly a constant, thus making step-wise increasing MSD. On the other hand, in this case
such a feature in the MSD does not occur because of the repitition of moving and stopping but continuous and periodic change 
from ultraslow ($\alpha(t)\sim0$) to ballistic diffusion ($\alpha(t)\sim1$).

The oscillating replication probability implies periodic and continuous alternation of the phase of the process between superdiffusion and subdiffusion.
It can be compared to a dynamic system of intermittent locomotion which 
have been importantly studied by intermittent search models
where discontinuous transition between explicitly defined two different diffusion phases are considered  \cite{is}.
Pauses, along with changes in the duration and speed of motion form different intermittent locomotions
which happens in contexts where animals adjust their motion to changing circumstances
and thus the cumulant distances over time show step-like picture as well as another oscillatory pictures with increasing and pausing intervals \cite{stepdistance}.

So we consider another oscillatory motion with the MSD which has increasing and pausing intervals,  
 \begin{figure}[ht]
 \includegraphics[width=8cm]{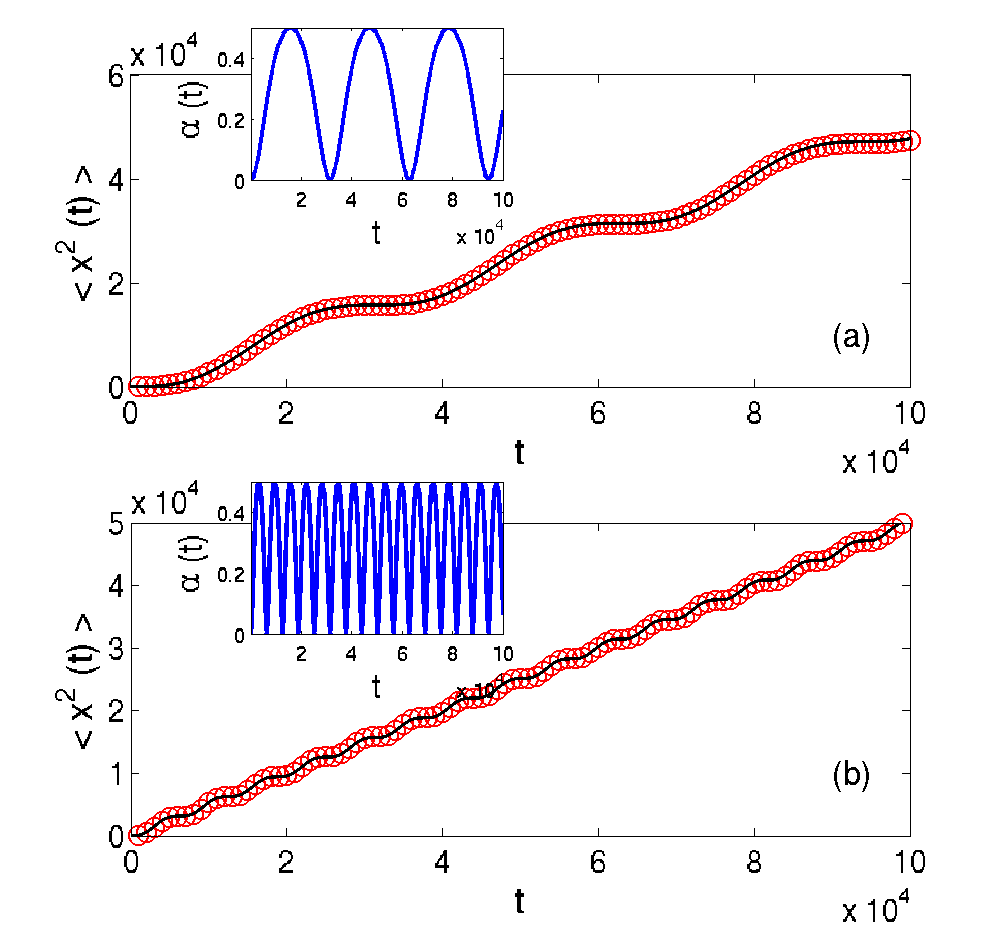}
 \caption{\label{fig2}Simulated MSDs of the model with $\alpha(t)$
obtained by $\langle x^2_t \rangle = t/2 - \text{sin} (2at)/4a +1/2 + \text{sin}2a/4a $ 
with (a) $a=10^{-5}$  and (b) $a=5 \times 10^{-5}$. An additive constant has been included
in the MSD here and after to meet the condition $\langle x^2_t \rangle = 1$ at time $t = 1$.
Circles representing the simulation data match well with the MSD functions (solid lines). 
Corresponding $\alpha(t)$'s are plotted in the insets. 
 }
 \end{figure}
$\langle x^2_t \rangle \sim t/2 - \text{sin} (2at)/4a$.
$a$ is a constant and the MSDs with $a=10^{-5}$ and  $a=5 \times 10^{-5}$
are shown in the Fig.~\ref{fig2} (a) and (b), respectively.
Corresponding $\alpha(t)$'s by Eq. (\ref{alpha}) with the MSDs oscillates
from 0 to 0.5 (the insets of Fig.~\ref{fig2}). The MSD is composed of periodically repeating plateaus 
and smoothly increasing regions, which correspond to the regions with the dominant anti-replication 
and the normal diffusive regions where $\alpha(t) \sim 1/2$, respectively. 
The shape of the MSD is similar to that of the Fig.~\ref{fig1}, but the sudden increases have been smoothen
due to the smaller maximum value of $\alpha(t)$.

We have shown that different diffusive phases can be alternated by a oscillatory replication probability,
which implies that it may be a generative mechanism to be able to describe changes of diffusion phase with time shown in
the kinetics of cells in external stimuli and intermittent locomotions of animals. 
However, what we have shown here is not about a specific system but about generic changes of diffusive phases,
and it needs to be more closely anaylzed to find which specific $\alpha(t)$ is appropriate to describe a specific system.

\subsection{Superdiffusion with $H=0.9$}

 \begin{figure}[ht]
 \includegraphics[width=9cm]{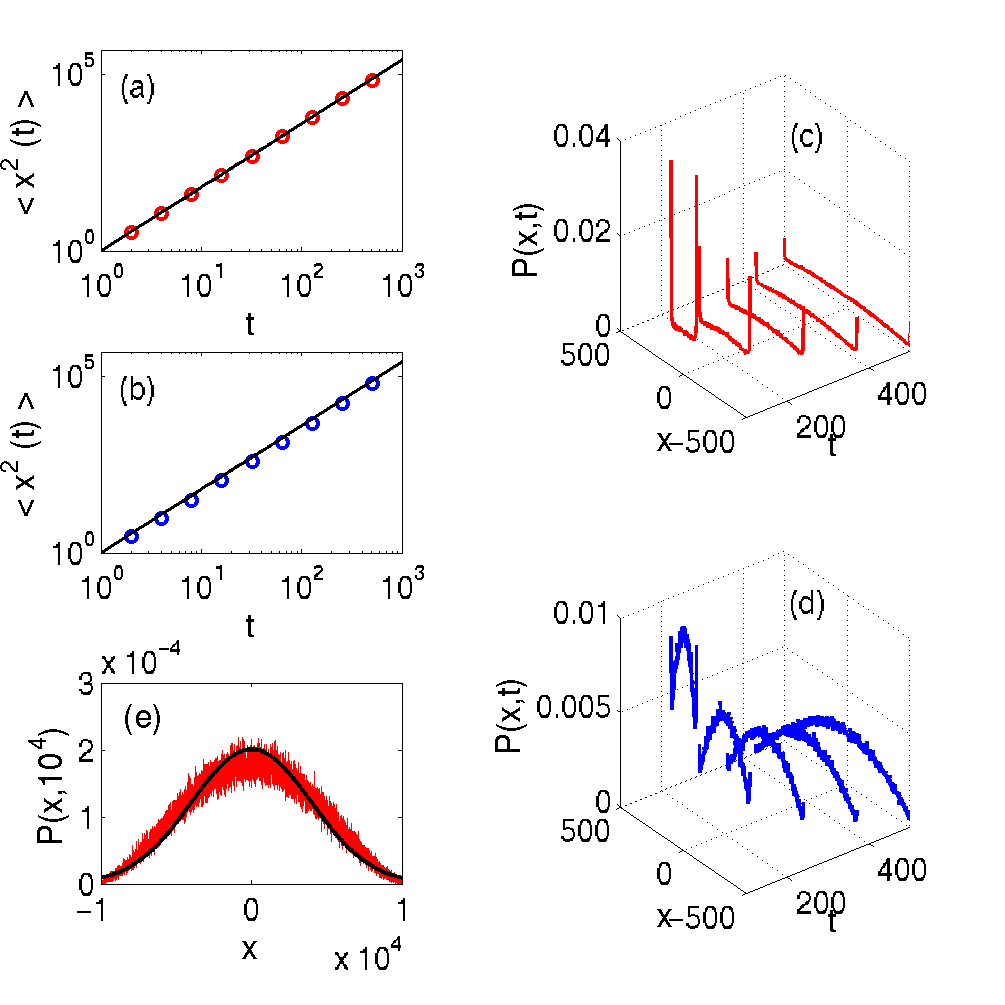}
 \caption{\label{fig3} Simulation results of models created by the replication probabilities related to
the given MSD $\langle x^2_t \rangle = t^{2H}$ with $H=0.9$.
(a) The MSD data obtained using the $\alpha(t)$ in Eq. (\ref{alpha}). The solid line represents the given MSD.
(b) The MSD data obtained using $\alpha_D (t)$ which ignores the second order derivative of the MSD in Eq.~(\ref{alpha}).
Results show that the models generate the MSD expected, 
while $\alpha(t)$ gives slightly more accurate coincidence witn the MSD than $\alpha_D (t)$.
Figure (c) and (d) show the PDFs corresponding to (a) and (b), respectively.
Shape of the PDF is strikingly different from the Gaussian, indicating
the effect of the second term on the left hand side of Eq.~(\ref{fpe}) at early times.
(e) The PDF at time $t=10^4$ for the model (a). The solid line represents the corresponding Gaussian
curve, which shows that the PDF  will eventually converge to the Gaussian distribution
after a sufficiently long time.
 }
 \end{figure}

\begin{figure}[ht]
 \includegraphics[width=8.5cm]{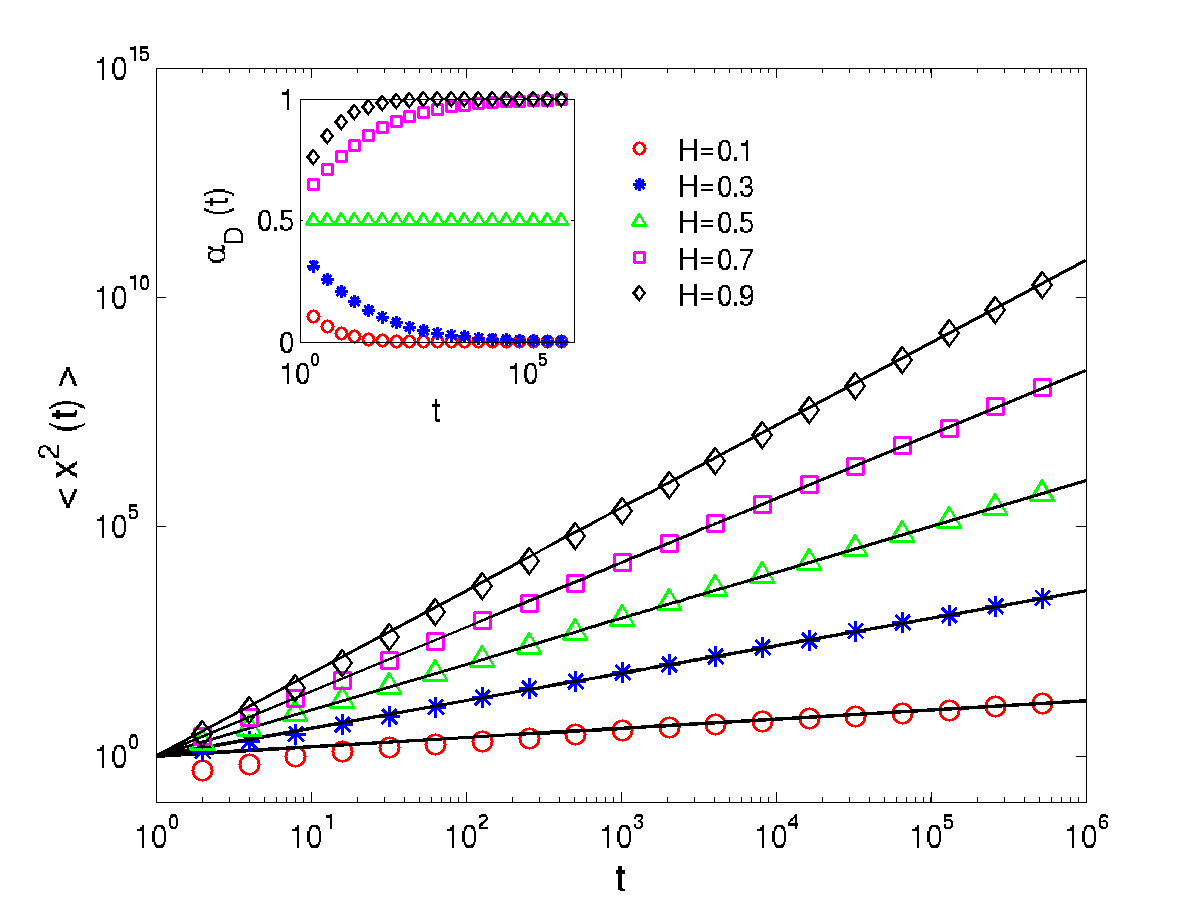}
 \caption{\label{fig4} Plots of the MSDs for anomalous diffusions with various $H$ 
induced by $\alpha_D (t)$ obtained from $\langle x^2_t \rangle = t^{2H}$.
In the inset, the corresponding $\alpha_D (t)$'s are plotted.
The symbols representing the simulation data fall excellently on the solid lines of $t^{2H}$.
 }
 \end{figure}

To compare the early behaviors of  two cases  where  $\alpha (t)$ of Eq. (\ref{alpha}) and 
$\alpha_D (t)$ obtained by ignoring $\ddot{\langle x^2_t \rangle}$ in Eq. (\ref{alpha})  are used, respectively,
we have used the MSD, $\langle x^2_t \rangle = t^{2H}$ with $H=0.9$. 
Fig. \ref{fig3} (a)  shows the MSD obtained through the simulation using the $\alpha (t)$,
which shows excellent agreement with the given MSD.
In Fig. \ref{fig3} (b), we also plotted the simulation result using $\alpha_D (t)$.
It shows that the data slightly deviate from the expected line.
Taking the second derivative in Eq.~(\ref{alpha}) into account gives more accurate MSD.
Although there is just a slight difference in the two MSDs, note that at the early times, $P(x,t)$ obtained 
by $\alpha(t)$ is distinguished from the Gaussian distribution showing the peaks at the possible maximum 
distances (Fig. \ref{fig3} (c)), while $P(x,t)$ obtained by $\alpha_D (t)$ relatively follows the Gaussian 
(Fig. \ref{fig3} (d)). However, at large times $P(x,t)$ obtained  by $\alpha (t)$ also converges 
to the Gaussian distribution (Fig. \ref{fig3} (e)) and thus it is sufficient to take just $\alpha_D (t)$ for the most of 
asymptotic behaviors (see Figure \ref{fig4}).
The peaks of $P(x,t)$ at early times
which resembles that of the L\'{e}vy walks \cite{levir}
is because the divergence of $\mathcal{R}(t)$ is not so fast 
enough to make the second order derivative term in $t$ in Eq. (\ref{fpe}) dominant, but significantly slows the convergence
of $P(x,t)$ to the Gaussian distribution. In such cases, the effect survives 
in early dynamics of stochastic processes.
Analysis of early dynamics is important in real and experimental environments
in which it is difficult to take a sufficiently long time and thus the second term on the left hand side of  Eq. (\ref{fpe}) could play a important role
in such contexts.
\subsection{Anomalous diffusions}

We have also considered anomalous diffusions induced by the NMRP model.
In Fig. \ref{fig4}, the MSDs for the anomalous diffusions
with the Hurst exponents ranging from 0.1 to 0.9 are shown. Models have been generated with
$\alpha_D (t)$ obtained by setting $\langle x^2_t \rangle = t^{2H}$. 
For $H>0.5$, $\alpha_D (t)$ increases with time, which induces the persistence in the process over time,
resulting in the superdiffusions.
While for $H<0.5$, $\alpha_D (t)$ decreases with time, which invokes the anti-persistence showing 
the subdiffusions.
Similar conclusion has been reported  using the latest memory enhancement model \cite{mm1}
which can be thought of as a special case of the NMRP model. 
For instance, the positive latest memory enhancement model can be reproduced 
in the framework of the NMRP
\begin{figure}[ht]
 \includegraphics[width=8cm]{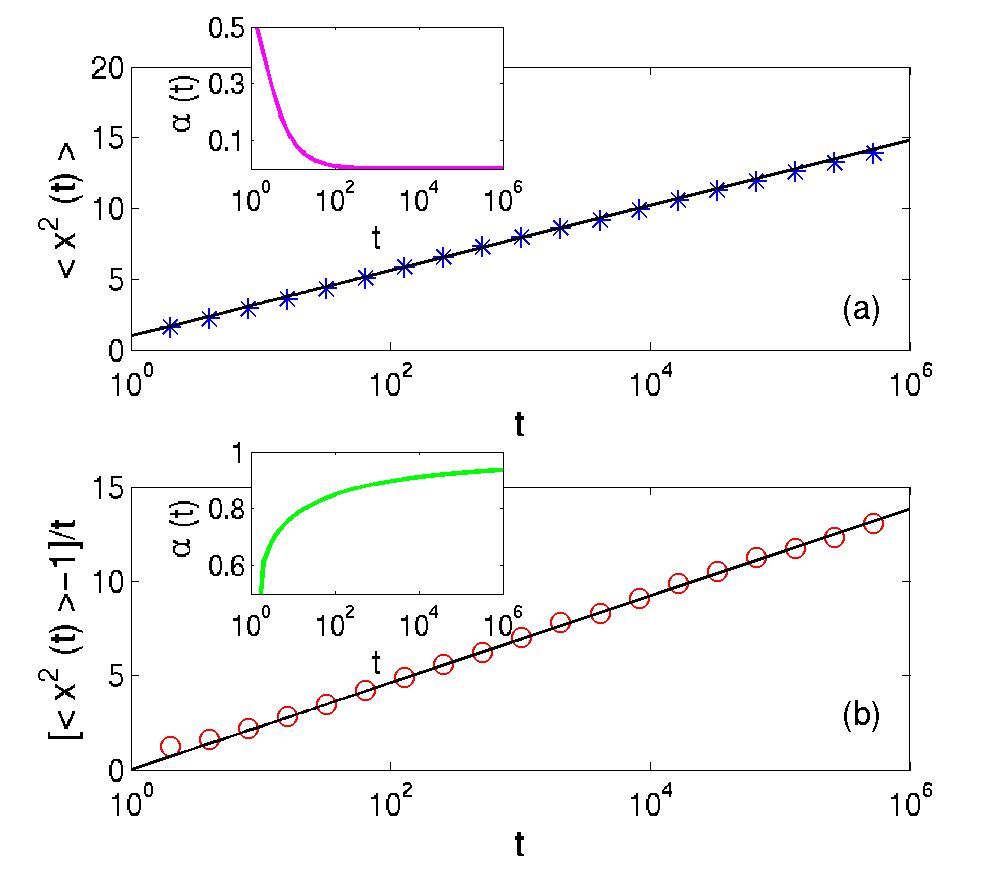}
 \caption{\label{fig5} The plot of the simulated MSDs using (a) $\langle x^2_t \rangle = \text{ln} t + 1$ and
(b)  $\langle x^2_t \rangle = t \text{ln} t +1$.
The insets shows the corresponding $\alpha (t)$ calculated from Eq.~(\ref{alpha}), respectively.
Stars and circles represent the data obtained from the simulation and the solid lines in each plots 
represent the analytic functions of the MSD.
}
 \end{figure}
if we use $\alpha(t)=1-1/2t^p$ with $p$ being the memory parameter.

\subsection{Marginal diffusions}

Fig. \ref{fig5} shows simulation results for logarithmic MSDs which are generated by $\alpha(t)$'s using
(a) $\langle x^2_t \rangle \sim \text{ln}t$, and (b) $\langle x^2_t \rangle \sim t\text{ln}t$.
In each figures, excellent agreements between the given MSDs and the simulation results are shown.
In the NMRP model, logarithmically slow subdiffusion is achieved by fast decreasing replication probability 
from the value of 0.5 (the inset of Fig. \ref{fig5} (a)), that is, the probability that the walker escapes away from a position  
decreases more rapidly than that of subdiffusions with time and the anti-persistence is strongly developed.
Such ultra-slow diffusions have been reported in various  contexts \cite{boyer,usbm,uctrw,sinai,us1,us2}, and
often arise as a marginal behavior of the subdiffusion with $H=0$.

The MSD of the type of $t\text{ln}t$ also appears in the marginal behaviors of the superdiffusions 
\cite {mm1, pre70, marginal}, while in the NMRP model it is embodied 
with $\alpha(t)$ increasing slowly compared to the cases of superdiffusions
as shown in the Fig.~\ref{fig5} (b). Thus the marginal behaviors of anomalous diffusions
can be also induced by a single origin, the NMRP with appropriate replication probabilities.

\section{Conclusion}
In conclusion, we have proposed a non-stationary random walk model
in which the steps are given by a Markov process replicating the immediate step with a time-varying
probability. The master equation for the probability
has been analytically acquired and the 
generalized TE for the PDF has been derived, from which we have obtained the general relation between the time-varying 
replication probability and the MSD with accuracy up to the second order. 
We realized several interesting cases such as  alternating diffusions, anomalous diffusions, 
and marginal diffusions.
Although the stepping process is Markovian, 
the time-varying nature of replication develops the long-term correlation between steps and 
the corresponding diffusive behaviors, i.e.
ballistic, super, sub, slow-sub,  and ultraslow diffusive phases as well as normal diffusion
have been induced depending on the values of the replication probability changing in time.
For oscillatory replication probability, alternating diffusions of different diffusion phases
have been induced, increasing (decreasing) replication probability with the value larger(smaller) than 0.5 
have caused superdiffusions (subdiffusions).
This single mechanism inducing diverse diffusions may provide a theoretical guide to 
experimental results of various types of diffusions
and furthermore, non-stationary stochastic processes shown in diverse fields. 
\\
We also remark that the further studies of the relation between a general replication probability 
and the autocorrelation function of steps will be helpful to elucidate the actual mechanism generating these long-term
correlations. Non-stationarity of the model would invoke the ergodicity breaking 
\cite{ergodicity,metzler} and the characteristics represented by the time average should be dealt  
separately from the results in this study which are obtained from the ensemble averages.

\end{document}